\documentstyle[nolta,epsf]{article}
\begin{document} 
\newcommand{\be}{\begin{equation}}
\newcommand{\ee}{\end{equation}}
\newcommand{\TS}[1]{{\tt <T>}{\bf #1}{\tt </T>}}
\newcommand{\AS}[1]{{\tt <A>}{\bf #1}{\tt </A>}}

\title{CONSTRAINED RANDOMISATION OF TIME SERIES\\ FOR HYPOTHESIS TESTING}

\author{Thomas Schreiber and Andreas Schmitz}
\affiliation{Physics Department, University of Wuppertal,
    D--42097 Wuppertal, Germany\\
    {\tt schreibe@theorie.physik.uni-wuppertal.de}}
\maketitle
\abstract
  We propose a general scheme to create time sequences that fulfill
  given constraints but are random otherwise. Significance levels for
  nonlinearity tests are as usually obtained by Monte Carlo
  resampling. In a new scheme, constraints including
  multivariate, nonlinear, and nonstationary properties are implemented in the
  form of a cost function.
\endabstract

\section{INTRODUCTION}
There are two distinct motivations to use a nonlinear approach when analysing
time series data. It might be that the arsenal of linear methods has been
exploited thoroughly but all the efforts left certain structures in the time
series unaccounted for. It is also common that a system is known to include
nonlinear components and therefore a linear description seems unsatisfactory in
the first place. The latter reasoning is rather dangerous since the
nonlinearity may not be reflected in a specific signal. In particular, we don't
know if it is of any practical use to go beyond the linear approximation.
Consequently, the application of nonlinear time series methods has to be
justified by establishing nonlinearity in the time series data.

This paper will discuss formal statistical tests for nonlinearity.  First, a
suitable null hypothesis for the underlying process will be formulated covering
all Gaussian linear processes and, in fact, a class that is somewhat wider.  We
will then test this null hypothesis by comparing a nonlinear parameter with its
probability distribution for the null hypothesis which has to be estimated by a
Monte Carlo resampling technique. This procedure is known in the nonlinear time
series literature as the method of {\em surrogate data}, see
Ref.~\cite{theiler1}. Thus we have to face a two-fold task. We have to find a
nonlinear parameter that is able to actually detect an existing deviation of
the data from a given null hypothesis and we have to provide an ensemble of
randomised time series that accurately represents the null hypothesis.

\section{DETECTING NONLINEARITY}

Several quantities have been discussed that can be used to characterise
nonlinear time series~\cite{power}. For the purpose of nonlinearity testing we
need such quantities that are particularly powerful in discriminating linear
dynamics and weakly nonlinear signatures. Traditional measures of nonlinearity
are derived from generalisations of the two-point autocovariance function or
the power spectrum. One particularly useful third order quantity is
$\sum_{n=\tau+1}^N (s_n-s_{n-\tau})^3$ since it measures the asymmetry of a
series under time reversal.  When a nonlinearity test is performed with the
question in mind if nonlinear deterministic modeling of the signal may be
useful, it seems most appropriate to use a test statistic that is related to a
nonlinear deterministic approach~\cite{ourbook}.  Widely used are test
statistics which in some way or the other quantify the nonlinear predictability
of the signal. Let $\vec x_n=(s_{n-(m-1)\tau},\ldots,s_n)$ be the sequence of
time delay embedding vectors obtained from the scalar time series $s_n$. The
nonlinear prediction error can then be defined by $\sqrt{\sum [\vec
x_{n+1}-F(\vec x_n)]^2}$. The prediction $F$ over one time step is performed by
averaging over the future values of all neighboring delay vectors $\vec x_{n'}$
closer to $\vec x_n$ than $\epsilon$ in $m$ embedding dimensions.

\section{SURROGATE DATA}
Almost all measures of nonlinearity have in common that their probability
distribution on finite data sets is not known analytically. It is therefore
necessary to use a Monte Carlo resampling technique. Traditional bootstrap
methods~\cite{efron} use explicit model equations that have to be extracted
from the data.  This {\em typical realizations} approach can be very powerful
for the computation of confidence intervals, provided the model equations can
be extracted successfully. As discussed by Theiler and Prichard~\cite{tp}, the
alternative approach of {\em constrained realizations} is more suitable for the
purpose of hypothesis testing we are interested in here. It avoids the fitting
of model equations by directly imposing the desired structures onto the
randomised time series.  However, the choice of possible null hypothesis is
limited by the difficulty to impose arbitrary structures on otherwise random
sequences. A general method has been recently proposed by one of the
authors~\cite{anneal}. 

It is essential for the validity of the statistical test that the surrogate
series are created properly. If they contain spurious differences to the
measured data, these may be detected by the test and interpreted as signatures
of nonlinearity. A simple case is the null hypothesis that the data consists of
independent draws from a fixed probability distribution. Surrogate time series
can be simply obtained by randomly shuffling the measured data. If we find
significantly different serial correlations in the data and the shuffles, we
can reject the hypothesis of independence.

\subsection{Fourier based methods}
A step towards more interesting null hypotheses is to incorporate the
structures reflected by linear two-point autocorrelations. A corresponding null
hypothesis is that the data have been generated by some linear stochastic
process with Gaussian increments. The statistical test is complicated by the
fact that we don't want to test against one particular linear process only (one
specific choice of ARMA coefficients), but against a whole class of
processes. This is called a {\em composite} null hypothesis.  The unknown
coefficients are sometimes referred to as {\em nuissance parameters}. There are
basically three directions we can take in this situation. First, we could try
to make the discriminating statistic independent of the nuissance
parameters. This approach has not been demonstrated to be viable for any but
some very simple statistics. Second, we could determine which linear model is
most likely realised in the data by a fit for the coefficients, and then test
against the hypothesis that the data has been generated by this particular
model. Surrogates are simply created by running the fitted model. The main
drawback is that we cannot recover the {\em true} underlying process by any fit
procedure.

The null hypothesis of an underlying Gaussian linear stochastic process can
also be formulated by stating that all structure to be found in a time series
is exhausted by computing first and second order quantities, the mean, the
variance and the autocovariance function.  This means that a randomised sample
can be obtained by creating sequences with the same second order properties as
the measured data, but which are otherwise random. When the linear properties
are specified by the squared amplitudes of the Fourier transform (that is, the
periodogram estimator of the power spectrum), surrogate time series are readily
created by multiplying the Fourier transform of the data by random phases and
then transforming back to the time domain.

The most obvious deviation from the Gaussian linear process is usually that the
data don't follow a Gaussian distribution. There is a simple generalisation of
the null hypothesis that explains deviations from the normal distribution by
the action of a monotone, static measurement function: $s_n=s(x_n)$ where
$\{x_n\}$ is a realisation of an ARMA process.  We want to regard a time series
from such a process as essentially linear since the only nonlinearity is
contained in the --- in principle invertible --- measurement function
$s(\cdot)$.

The most commonly used method to create surrogate data sets for this null
hypothesis essentially attempts to invert $s(\cdot)$ by rescaling the time
series $\{s_n\}$ to conform with a Gaussian distribution. The rescaled version
is then phase randomised (conserving Gaussianity on average) and the result is
rescaled to the empirical distribution of $\{s_n\}$. Schreiber and
Schmitz~\cite{surrowe} argue that for short and strongly correlated sequences,
this algorithm can yield an incorrect test due to a bias towards a flat
spectrum. They propose a method which iteratively corrects deviations in
spectrum and distribution.  Alternatingly, the surrogate is filtered towards
the correct Fourier amplitudes and rank-ordered to the correct
distribution. The accuracy that can be reached depends on the size and
structure of the data and is generally sufficient for hypothesis testing.

\subsection{General scheme}
The above schemes are based on the Fourier amplitudes of the data which is
appropriate in many cases. However, there remain some flaw, the strongest being
the severely restricted class of testable null hypotheses.  In the general
approach of Ref.~\cite{anneal}, constraints ({\sl e.g.}  autocorrelations) on
the surrogate data are implemented by a cost function $E(\{s_n\})$ which has a
global minimum when the constraints are fulfilled. This cost function will be
minimised by simulated annealing~\cite{annealbook}. Starting with a random
permutation of the original time series, the surrogate is modified by
exchanging two points chosen at random. The modification will be accepted if it
yields a lower value for the cost function or else with a probability
$p=\exp(-\Delta E/T)$. The ``system temperature'' $T$ will be lowered slowly to
let the system settle down to a minimum.

The constraint that the autocovariances of the surrogate $C'(\tau)$ should be
the same as those of the data $C(\tau)$ can be realised by specifying the
discrepancy as a cost function, for example
\be
\label{eq:cost}
 E=\sum_{\tau=0}^{N-1} |C'(\tau)-C(\tau)| 
\,.\ee
Now $E(\{{\tilde s}_n\})$ is minimised among all permutations $\{{\tilde
s}_n\}$ of the original time series $\{s_n\}$. With an appropriate cooling 
scheme, the annealing procedure can reach any desired accuracy.  

Constrained randomisation using combinatorial minimisation is a very flexible
method since in principle arbitrary constraints can be realised. It can be
quite useful to be able to incorporate into the surrogates any feature of the
data that is understood already or that is considered to be uninteresting.  The
price for the accuracy and generality of the method is its high computational
cost.

\section{APPLICATIONS}

\begin{figure}
   \centerline{
% GNUPLOT: LaTeX picture with Postscript
\setlength{\unitlength}{0.1bp}
\begin{picture}(1980,1511)(0,0)
\includegraphics{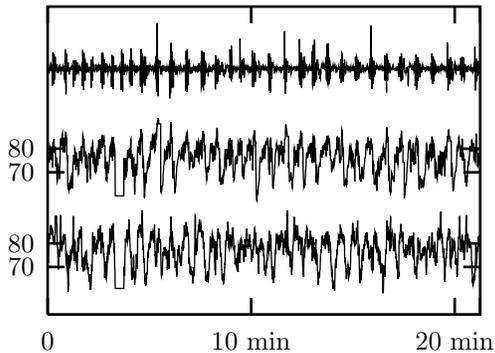}
\put(1834,150){\makebox(0,0){20 min}}
\put(1067,150){\makebox(0,0){10 min}}
\put(300,150){\makebox(0,0){0}}
\put(250,429){\makebox(0,0)[r]{70}}
\put(250,518){\makebox(0,0)[r]{80}}
\put(250,786){\makebox(0,0)[r]{70}}
\put(250,875){\makebox(0,0)[r]{80}}
\end{picture}
}\vspace{-10pt}
   \caption[]{\label{fig:multi} Simultaneous measurements of breath and heart
   rates, upper and middle traces. Lower trace: a surrogate heart
   rate series preserving the autocorrelation structure and the
   cross-correlation to the fixed breath rate series, as well as a spurious 
   gap in the
   data. Auto- and cross-correlation together seems to explain some, but not
   all of the structure present in the heart rate series.}
\end{figure}

\hspace*{\parindent}{\boldmath $Heart$ $rate$}{\bf---} Let us give an example
for the flexibility of the approach, a simultaneous recording of the breath
rate and the instantaneous heart rate of a human subject during
sleep~\cite{gold}, see Fig.~\ref{fig:multi}. An interesting question is, how
much of the structure in the heart rate (middle) can be explained by linear
dependence on the breath rate (upper). In order to answer this question, we
need to make surrogates that have the same autocorrelation structure but also
the same cross-correlation with respect to the fixed input signal, the breath
rate. Accordingly, a constraint is used to fix the auto-covariance and the
cross-covariance with the reference (breath) signal.  While the linear
cross-correlation with the breath rate explains the cyclic structure of the
heart rate data, other features, in particular the asymmetry under time
reversal, remain unexplained. Possible explanations include artefacts due to
the peculiar way of deriving heart rate from inter-beat intervals, nonlinear
coupling to the breath activity, nonlinearity in the cardiac system, and
others.

\begin{figure}
   \centerline{
% GNUPLOT: LaTeX picture with Postscript
\setlength{\unitlength}{0.1bp}
\begin{picture}(1980,1511)(0,0)
\includegraphics{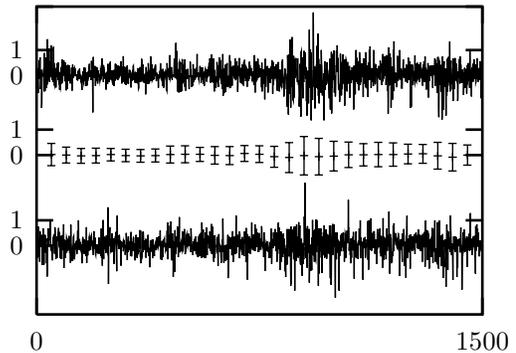}
\put(1930,150){\makebox(0,0){1500}}
\put(250,150){\makebox(0,0){0}}
\put(200,1247){\makebox(0,0)[r]{1}}
\put(200,1151){\makebox(0,0)[r]{0}}
\put(200,947){\makebox(0,0)[r]{1}}
\put(200,851){\makebox(0,0)[r]{0}}
\put(200,605){\makebox(0,0)[r]{1}}
\put(200,510){\makebox(0,0)[r]{0}}
\end{picture}
}\vspace{-10pt}
   \caption[]{Nonstationary financial time series (top)
      and a surrogate (bottom) preserving the nonstationary structure
      quantified by running window estimates of the local mean and variance
      (middle).\label{fig:bund}}
\end{figure}

{\boldmath $Financial$ $data$}{\bf ---} Let us study 1500 daily returns (until
the end of 1996) of the {\em BUND Future}, a derived financial instrument of
the German stock market. (Data by courtesy of Thomas Sch\"urmann, WGZ-Bank
D\"usseldorf.)  The sequence (Fig.~\ref{fig:bund}, upper) is nonstationary in
the sense that the local mean and variance undergo changes on a
time scale that is long compared to the fluctuations of the series itself. This
property is known in the statistical literature as {\em heteroscedasticity} and
modeled by the so-called GARCH and related models. Here, we want to avoid the
construction of a parametric model but rather ask the question if
the data is compatible with the null hypothesis of a correlated linear
stochastic process with time dependent local mean and variance. We can answer
this question in a statistical sense by creating surrogate time series that
show the same linear correlations and the same time dependence of the running
mean and variance as the data and comparing a nonlinear statistic
between data and surrogates. Accordingly, a cost function is set up to match
the autocorrelation function up to five days and the moving mean and variance
in sliding windows of 100 days duration. Using a time-asymmetry statistic, the
null hypothesis could not be rejected, suggesting that the above
characterisation of the data is operationally complete.

{\boldmath $Unevenly$ $sampled$ $data$}{\bf ---} 
Let us finally show how the new randomisation method can be used to test for
nonlinearity in time series with time intervals of different sizes. Unevenly
sampled data are quite common, examples include drill core data, astronomical
observations or stock price notations. Most observables and algorithms cannot
easily be generalised to this case which is particularly true for nonlinear time
series methods.  Interpolating the data to equally spaced sampling times is
not recommendable for a test for nonlinearity since one could not {\sl a
posteriori} distinguish
between genuine structure and nonlinearity introduced spuriously by the
interpolation process.

\begin{figure}
\centerline{\epsffile{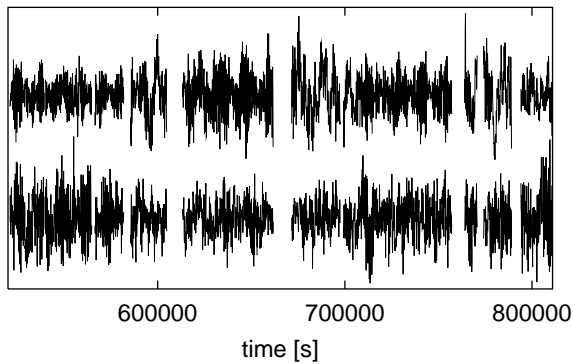}}\vspace{-10pt}
\caption[]{\label{fig:esur}
  The down-sampled data set E with one corresponding surrogate. Gaps
  of different sizes prevents reasonable interpolation.  }
\end{figure}

Consider a time series sampled at times $\{t_n\}$ that need not be equally
spaced. The power spectrum can then be estimated by the Lomb periodogram
$P(\omega)$, as discussed for example in Ref.~\cite{NR_lomb}.  For time series
sampled at constant time intervals, the Lomb periodogram yields the standard
squared Fourier transformation.  Except for this particular case, it does not
have any inverse transformation, which makes it impossible to use the standard
surrogate data algorithms mentioned above. Therefore, we use the Lomb
periodogram of the data as a constraint for the surrogates.  It can be
expressed as a cost function for example by: $E=\sum_{k=1}^{N_f}
|P'(k\omega_0)-P(k\omega_0)|$.  We use $P$ at $N_f$ equally spaced frequencies
$k\omega_0$, other choices are possible.  Consider a series~\cite{set_e} of the
time-integrated intensity of light observed from a variable star, see
Fig.~\ref{fig:esur}. It consists of 17 parts with different numbers of points,
the time range of which may overlap or show gaps of up to 10000~s. Between
gaps, the (down-sampled) data is evenly sampled with $\Delta=120$~s, the total
number of points is $N=2260$.  The linear null hypothesis was not rejected by
the time reversibility statistic. One surrogate is shown in
Fig.~\ref{fig:esur}.

\section{DISCUSSION}
We have set up a statistical hypothesis test of nonlinearity. How interesting
its outcome is depends on the specific null hypothesis chosen. The most
meaningful test can be performed if the null hypothesis is plausible enough so
that we are prepared to believe it in the lack of evidence against it.  In
general, we will specify a set of observables we believe to be complete to
describe the structure found in the data. The surrogates will then share these
properties with the data and any significant discrepancy between data and
surrogates can guide to a more complete understanding.

Recent efforts on the generalisation of randomisation schemes try to
broaden the repertoire of null hypothesis we can test against. The
hope is that we can eventually choose one that is general enough to be
acceptable if we fail to reject it with the methods we have. Still, we
cannot prove that there isn't any structure in the data beyond what is
covered by the null hypothesis.  From a practical point of view,
however, there is not much of a difference between structure that
isn't there and structure that is undetectable with our observational
means.

We like to thank Daniel Kaplan, James Theiler, Peter Grassberger and
Holger Kantz for useful discussions. This work was supported by the
SFB 237 of the Deutsche Forschungsgemeinschaft.

\end{document}